\documentclass[twocolumn,pre,floatfix]{revtex4}

\usepackage[dvips]{graphicx}
\usepackage[dvips]{color}
\usepackage{amsmath}
\usepackage{amssymb}
\usepackage{bm}
\usepackage{times}
\usepackage{textcomp}

\newcommand{\vct}[1]{\mathbf{#1}}
\newcommand{\uvct}[1]{\mathbf{\hat{#1}}}

\newcommand{\grad}{\bm{\nabla}}

\newcommand{\rhodna}{\rho_{\text{DNA}}}

\begin{document}

\title{Colloids dragged through a polymer solution: experiment, theory and
simulation}
\date{\today}

\author{Christof Gutsche}
\email{gutsche@physik.uni-leipzig.de}
\affiliation{Institut f{\"u}r Experimentalphysik~I, Universit{\"a}t Leipzig, 04103 Leipzig, Germany}
\author{Friedrich Kremer}
\email{kremer@physik.uni-leipzig.de}
\affiliation{Institut f{\"u}r Experimentalphysik~I, Universit{\"a}t Leipzig, 04103 Leipzig, Germany}
\author{Matthias Kr{\"u}ger}
\email{matthias.krueger@uni-konstanz.de}
\altaffiliation[Current address: ]{Fachbereich Physik, Universit{\"a}t
Konstanz, D-78457 Konstanz, Germany}
\affiliation{Max-Planck-Institut f{\"u}r
Metallforschung, Heisenbergstr.\ 3, 70569 Stuttgart, Germany, and \\
Institut f{\"u}r Theoretische und Angewandte Physik, Universit{\"a}t 
Stuttgart,  Pfaffenwaldring 57, 70569 Stuttgart, Germany}
\author{Markus Rauscher}
\affiliation{Max-Planck-Institut f{\"u}r
Metallforschung, Heisenbergstr.\ 3, 70569 Stuttgart, Germany, and \\
Institut f{\"u}r Theoretische und Angewandte Physik, Universit{\"a}t 
Stuttgart,  Pfaffenwaldring 57, 70569 Stuttgart, Germany}
\author{Rudolf Weeber}
\affiliation{Institut f{\"u}r Computerphysik, Universit{\"a}t Stuttgart, Pfaffenwaldring 27,
70569 Stuttgart, Germany}
\author{Jens Harting}
\affiliation{Institut f{\"u}r Computerphysik, Universit{\"a}t Stuttgart, Pfaffenwaldring 27,
70569 Stuttgart, Germany}

\begin{abstract}
We present micro-rheological measurements of the drag force on
colloids pulled through a solution of $\lambda$-DNA (used here as a
monodisperse model polymer) with an optical tweezer. The
experiments show a violation of the Stokes-Einstein relation based
on the independently measured viscosity of the DNA solution: the
drag force is larger than expected. We attribute this to the
accumulation of DNA infront of the colloid and the reduced DNA
density behind the colloid. This hypothesis is corroborated by a
simple 
drift-diffusion model for the DNA molecules, which reproduces the
experimental data surprisingly well, as well as by corresponding
Brownian dynamics simulations.
\end{abstract}


\maketitle

\section{Introduction}\label{sec:intro}
Complex fluids in general and colloid-polymer mixtures in particular are
an ideal model system for studying of the structure and phase behaviour of
multicomponent systems.  But they also play a large role in many
technological processes such as oil recovery, food science
\cite{mezzenga05}, as well as in most biological systems. For these
systems, in addition to the equilibrium properties the dynamics is
important. While until recently, research largely focused on bulk
rheological properties, it is now understood that interface effects play a
significant role. The structure of fluids in general changes in the
vicinity of interfaces. In equilibrium situations this is known to lead to
fluid structure mediated interactions, i.e., the so-called solvation force
or depletion interactions. Theoretically, the concept of depletion
interactions in colloid-polymer solutions has been extended to
non-equilibrium situations \cite{dzubiella03b,krueger07}\/. While in
equilibrium, these interactions are short ranged (with an interaction
range on the order of the particle diameter), they become long ranged in
non-equilibrium situations. The reason for this are the long ranged
structural changes in the fluid. These structural changes, i.e., an
enhanced polymer density infront of the colloid, and a reduced polymer
density in its back, also lead to an enhanced friction
\cite{squires05b,rauscher07b}\/.  In addition, the structural changes in
the vicinity of walls couple back to the rheological properties there,
e.g., leading to slip boundary conditions \cite{tuinier05,tuinier06}\/.
This, on the other hand, leads to a drag reduction as compared to the
expected Stokes drag for the bulk fluid. This work was inspired by dynamic
light scattering experiments on colloids in solutions of macro molecules
\cite{ullmann85,phillies85,phillies87}\/.

In order to elucidate the dynamics of colloids in polymer solutions in
more detail, we perform highly controlled experiments on single isolated
colloids in $\lambda$-DNA solutions using optical tweezers. This technique
allows us to move a colloid through a highly mono-disperse polymer
solution at a given velocity and to measure the drag force on the colloid
with pN-resolution at the same time. For high DNA concentrations we find a
significantly higher drag force than predicted by the Stokes-Einstein
theory for the homogeneous solution, however, the force scales
approximately linearly with the velocity. We compare the experimental
results to a simplified dynamic density functional theory (DDFT) for the
non-interacting polymers in a flowing solvent  and to Brownian dynamics
(BD) simulations.

\section{Experiment}
\subsection{Materials and methods}
\begin{figure}
\includegraphics[width=\linewidth]{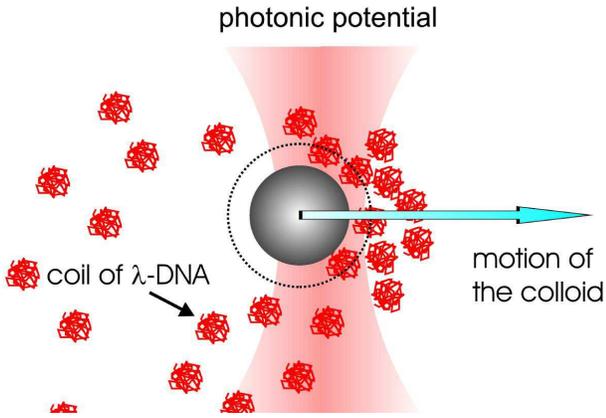}
\caption{\label{fig:exp} Illustration of the experimental setup.
The colloid (grey) sourrounded by coils of DNA (red) is hold in an
optical trap established by a photonic potential (magenta).}
\end{figure}

Fig.~\ref{fig:exp} illustrates the experimental set up, in which one
colloid is held by an optical trap surrounded by a polymer solution of
$\lambda$-DNA (obtained by New England BioLabs, Germany). An inverted
microscope (Axiovert S 100 TV, Carl Zeiss, Jena, Germany) is used and the
optical trap is realized with a diode pumped Nd:-YAG laser (1064~nm,
1~Watt, LCS-DTL 322; Laser 2000, Wessling, Germany).  Its power is
stabilized to achieve long-term stability.  Additionally, the profile of
the laser-beam was monitored. After passing an optical isolator, a
quarter-wave plate is used in order to produce circularly polarized light
to exclude effects due to reflection differences of the mirrors between
the p- and s-polarisation of the laser light. The beam is expanded and
coupled into the back aperture of the microscope objective (Plan-Neofluor
100$\times$1.30 Oil, Carl Zeiss, Jena, Germany). Video imaging and the
optical position detection are accomplished by a digital camera (KPF 120,
Hitachi, D{\"u}sseldorf, Germany). The optical stage is positioned in
three dimensions with nanometer resolution using piezoactuators (P-5173CD,
Physik Instrumente, Karlsruhe, Germany). The sample cell consists of a
closed chamber that can be flushed by a syringe pump with varying
solutions. The viscosities are measured with an Ostwald viscosimeter at
the same temperature (25\textdegree C room temperature) and
with the same $\lambda$-DNA solutions as used in the experiment.
\subsection{Data analysis and calibration}
\begin{figure}
\includegraphics[width=\linewidth]{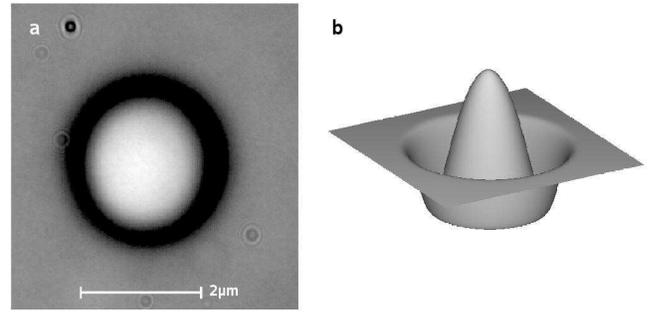}
\caption{\label{fig:image} (a) microscope image of a single colloid
($1.12 \mu$m radius) in the optical trap, (b) fit of the intensity
distribution for the image shown in (a) using the
Levenberg-Marquardt algorithm.}
\end{figure}

The position of the bead in the optical trap is determined by image
analysis (Fig.~\ref{fig:image}a) \cite{ovryn00a,ovryn00b}\/.  For that a
sequence of images is recorded (repetition rate 30~Hz) and analysed based
on the Levenberg-Marquardt algorithm (Fig.~\ref{fig:image}b). As fit
function 
\begin{eqnarray} 
I&=&I_0+A\,(1-a\,d)\,\exp(-d^3)\\
d&=&\left(\frac{x-x_0}{r}\right)^2+\left(\frac{y-y_0}{r}\right)^2
\end{eqnarray}
is used, where $(x_0, y_0)$ is the center position, $r$ the
optical radius, $A$ the amplitude of the profile relative to the
background image intensity $I_0$ and $a$ a constant to consider the dark
diffuse ring around the colloid.  A variation of $r$, $A$ or $a$ in an
image sequence indicates the motion in $z$-direction.  The calibration of
the optical trap is based on Stokes' law for the pure solvent (here
water) as described in detail elsewhere \cite{svoboda94}\/. A typical
force constant of the trap is 0.085~pN/nm corresponding to forces in the
range between 0--50~pN, which can be determined with an accuracy of
$\pm$0.15~pN\/. 

In the experiments, no sign of irreversible adsorption of $\lambda$-DNA
molecules to the colloid was detected. 
While switching between different flow velocities, the same forces at the
same speed were obtained in the range of our uncertainties and after
flushing back to pure water the Stokes force of the colloid could be
reproduced. 

\subsection{Experimental results}

\begin{figure}
\includegraphics[width=\linewidth]{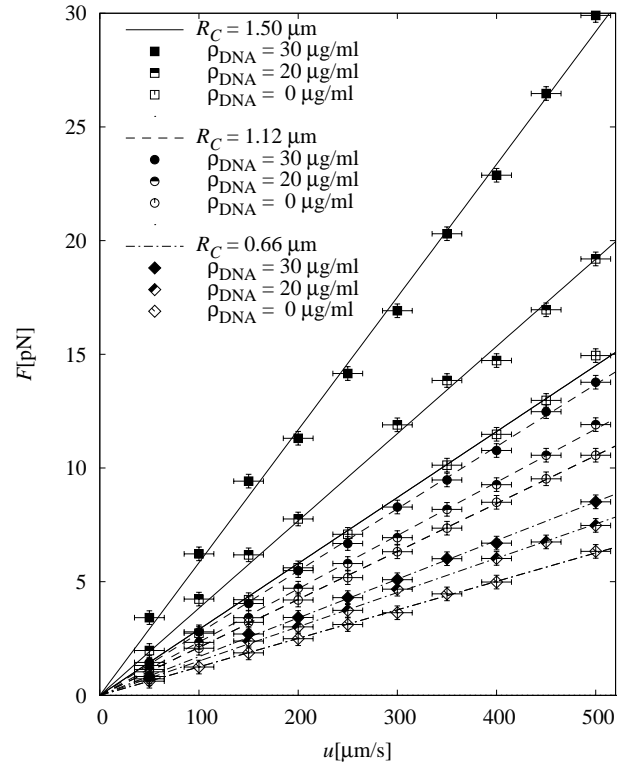}
\caption{\label{fig:radii} Drag force $F$ on colloids of radii
$R_C=1.5\,\mu$m (boxes), $1.12\,\mu$m (circles), and $0.66\,\mu$m
(diamonds) in $\rhodna=30\,\mu\text{g}/\text{ml}$ (full symbols) and
$20\,\mu\text{g}/\text{ml}$ (half filled symbols), as well as in pure
water (open symbols) as a function of the pulling speed $u$\/. Also shown
are linear fits to the data from which one can extract a density dependent
viscosity via the Stokes-Equation~(\protect\ref{eqstokes})\/. However, as
shown in Fig.~\protect\ref{viscofig}, these viscosities are inconsistent
as they depend on the colloid radius.
}
\end{figure}

\begin{table}
\begin{tabular}{|r||r|r|r|}\hline
$\rhodna=$ & 0~$\mu\text{g}/\text{ml}$ &
20~$\mu\text{g}/\text{ml}$ & 30~$\mu\text{g}/\text{ml}$\\ \hline\hline
$R_C = 1.50\,\mu\text{m}$ & 1.03~mPas & 1.36~mPas & 2.07~mPas \\\hline
$R_C = 1.12\,\mu\text{m}$ & 1.00~mPas & 1.12~mPas & 1.30~mPas \\\hline 
$R_C = 0.66\,\mu\text{m}$ & 1.01~mPas & 1.21~mPas & 1.37~mPas \\\hline
viscosimeter & 1.00~mPas & 1.09~mPas & 1.14~mPas\\\hline
\end{tabular}
\caption{\label{viscotab} Viscosity as a function of DNA concentration
extracted from linear fits to the data shown in
Fig.~\protect\ref{fig:radii} using the Stokes formula
Eq.~(\protect\ref{eqstokes}) compared to the viscosities measured in a
viscosimeter\/. From the viscosimeter data one obtains an intrinsic
viscosity of the DNA of $[\eta]=1.55$\/. See also
Fig.~\protect\ref{viscofig}\/.}
\end{table}

\begin{figure}
\includegraphics[width=\linewidth]{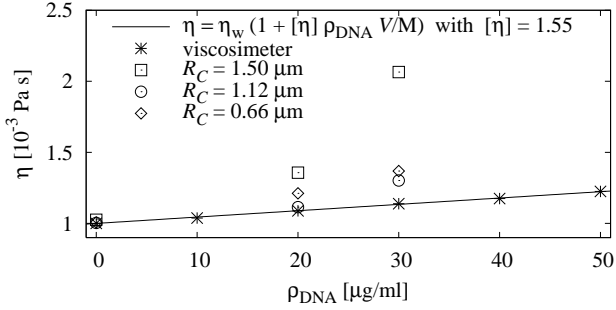}
\caption{\label{viscofig} Viscosity as a function of DNA
concentration extracted from linear fits to the data shown in
Fig.~\protect\ref{fig:radii} using the Stokes formula
Eq.~(\protect\ref{eqstokes}) compared to the viscosities measured
in a viscosimeter, c.f., Table~\protect\ref{viscotab}\/.
From the viscosimeter data one obtains an
intrinsic viscosity of the DNA of $[\eta]=1.55$\/.}
\end{figure}

\begin{figure}
\includegraphics[width=\linewidth]{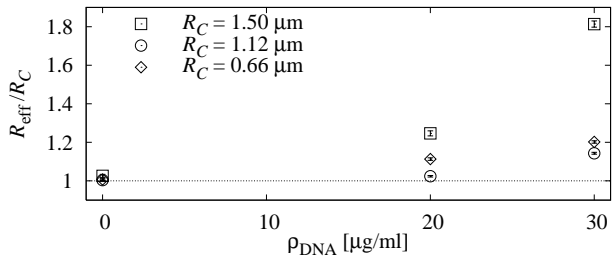}
\caption{\label{figeffrad} Effective colloid radius
$R_{\text{eff}}$ extracted from the linear fits to the data in
Fig.~\protect\ref{fig:radii} relative to the nominal radius
$R_C$ as a function of the DNA density $\rhodna$\/.}
\end{figure}

\begin{table}
\begin{tabular}{|r||r|r|r|}\hline
$\rhodna=$ & 0~$\mu\text{g}/\text{ml}$ & 20~$\mu\text{g}/\text{ml}$
& 30~$\mu\text{g}/\text{ml}$\\ \hline\hline
$R_C = 1.50\,\mu\text{m}$ & $1.53\,\mu\text{m}$ &
$1.87\,\mu\text{m}$ & $2.72\,\mu\text{m}$ \\\hline
$R_C = 1.12\,\mu\text{m}$ & $1.12\,\mu\text{m}$ &
$1.14\,\mu\text{m}$& $1.27\,\mu\text{m}$\\\hline 
$R_C = 0.66\,\mu\text{m}$ & $0.66\,\mu\text{m}$ &
$0.73\,\mu\text{m}$ & $0.79\,\mu\text{m}$\\\hline
\end{tabular}
\caption{\label{tabeffrad}Effective colloid radius $R_{\text{eff}}$
extracted from the linear fits to the data in
Fig.~\protect\ref{fig:radii}\/.}
\end{table}

\begin{figure}
\includegraphics[width=\linewidth]{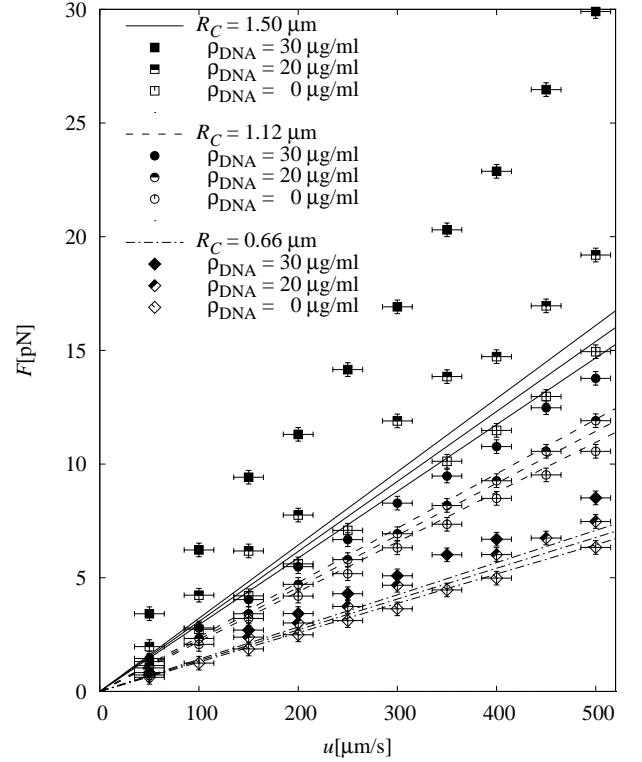}
\caption{\label{fig:radii_2} Drag force $F$ on colloids of radii
$R_C=1.5\,\mu$m (boxes), $1.12\,\mu$m (circles), and $0.66\,\mu$m
(diamonds) in
$\rho_{\rm DNA}=30\,\mu\text{g}/\text{l}$ (full symbols) and
$20\,\mu\text{g}/\text{l}$ (half filled symbols),
as well as in pure water (open symbols) as a function of the pulling 
speed $u$\/. Also shown is the Stokes force $F_S$ on the colloids as
expected for the viscosities measured for the given DNA
concentration in a viscosimeter, c.f., Table~\protect\ref{viscotab}
and Fig.~\protect\ref{viscofig}\/.}
\end{figure}

\begin{figure}
\includegraphics[width=\linewidth]{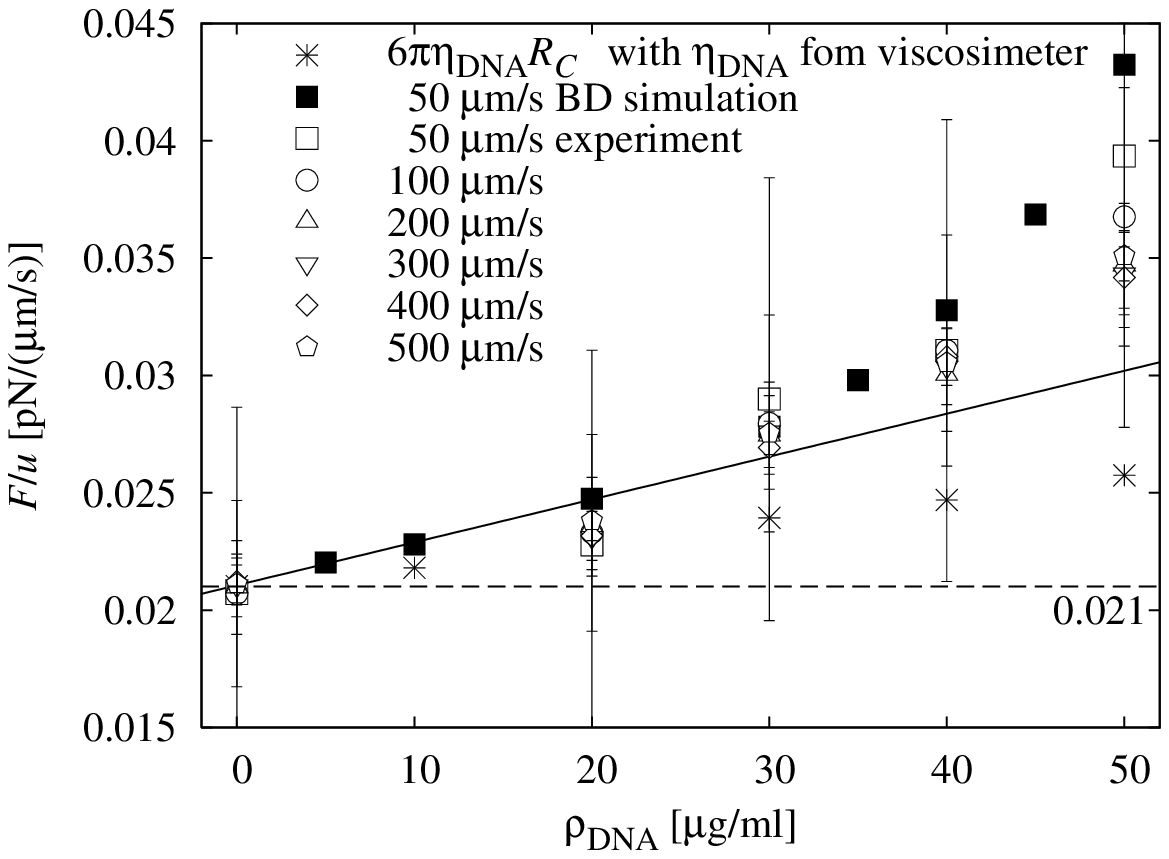}
\caption{\label{fig:collaps} The measured drag force $F$ on a colloid of radius
$1.12\,\mu$m  normalized to the velocity $u$ as a function of the DNA
concentration (open symbols)\/. For the clarity of the presentation only a subset of the experimental data is shown. 
The data collapse demonstrates that $F$ is
proportional to $u$, however, the dependence on $\rho_{\rm dna}$ is
non-linear. The drag is also significantly larger than
expected from the increased viscosity as measured in a
viscosimeter ({\large $\ast$})\/. In pure water we obtain
$F/u=0.021\,\text{pN}/(\mu \text{m}/\text{s})$ (dashed line)\/.
Also shown are simulation results for polymers with modified 
mobility (as explained in the text---full symbols)\/. 
A fit for concentrations between $0$ and $20\mu$g/ml 
(full line) highlights the nonlinearity in the density\/.}
\end{figure}

Fig.~\ref{fig:radii} shows the mesured force on colloids of sizes
$R_C=1.5\,\mu\text{m}$, $1.12\,\mu\text{m}$, and
$0.66\,\mu\text{m}$ as a function of the dragging velocity $u$ in
pure water as well as for
DNA concentrations $\rhodna=20\,\mu\text{g}/\text{ml}$ and
$\rhodna=30\,\mu\text{g}/\text{ml}$\/. Linear fits to the data 
analyzed in terms of the Stokes drag formula for a colloid of
diameter $R_C$ in a DNA solution of viscosity $\eta$
\begin{equation}
\label{eqstokes}
F_S^P = 6\,\pi\,\eta\,R_C\,u
\end{equation}
yield the DNA concentration dependent viscosities in
Table~\ref{viscotab}, shown in Fig.~\ref{viscofig}\/. The
viscosities obtained in this manner are significantly larger than
the viscosities measured in a viscosimeter for the same DNA
solution and they increase with the colloid radius $R_C$\/.
This is a strong indication that on top of the Stokes drag a second
mechanism plays a role.

Assuming the viscosities measured in the viscosimeter we can also
extract an effective hydrodynamic radius $R_{\text{eff}}$ from a
fit of Eq.~(\ref{eqstokes}) to the data in Fig.~\ref{fig:radii}\/.
The resulting $R_{\text{eff}}$ as a function of $\rhodna$
normalized to the nominal colloid radius $R_C$ is shown in
Table~\ref{tabeffrad} and Fig.~\ref{figeffrad}\/. $R_{\text{eff}}/R_C$ increases with
$\rhodna$ but it also varies with $R_C$: the value for
$R_C=1.5\,\mu$m is much larger than the values for the smaller
colloids. If the increase of
$R_{\text{eff}}$ was due to adsorption of DNA molecules to the
colloid we would expect the absolute increase of $R_{\text{eff}}$
to a first approximation to be independent of $R_C$\/. 

The molecular weight of a $\lambda$-DNA
molecule is $M=31.5\times 10^6\,\text{amu}=5.23\times 10^{-11}\,\mu\text{g}$
and its contour length is about $16\,\mu\text{m}$, leading to a
radius of gyration of roughly $R_g=0.5\,\mu\text{m}$\/. This
corresponds roughly to a hydrodynamic radius of $R_H= 0.662\,R_g =
0.33\,\mu\text{m}$\/. Based on the hydrodynamic radius and the
radius of gyration we get molecular volumes $V_H=1.5\times
10^{-13}\,\text{ml}$ and $V_g=5.2\times 10^{-13}\,\text{ml}$,
respectively. For the highest DNA concentrations used in the
experiment, i.e., $\rhodna= 50\,\mu\text{g}/\text{ml}$ this
corresponds to volume packing fractions of 14\%\/ and 50\%\/,
respectively. With this, we can extract the intrinsic viscosity
of the DNA solution from the viscosimeter data and get
$[\eta]=1.55$\/. Assuming the intrinsic viscosity for hard sphere
suspensions $[\eta]=5/2$ as predicted in \cite{einstein11,ford60} 
we get an effective hydrodynamic radius of $R_\eta =
0.28\,\mu\text{m}$\/.

In Fig.~\ref{fig:radii_2} we compare the measured drag forces on the
colloids with the predictions given by the Stokes equation
(\ref{eqstokes}), assuming the viscosity values measured
independently in the viscosimeter. The data is normalized to the
measurement in pure water. Apparently, in the DNA solution the drag
force is much larger than expected for the increase in viscosity.
In addition, the difference of the measured force to the prediction
assuming Stokes' equation (\ref{eqstokes}) with the viscosity
values measured in the viscosimeter is larger for larger colloids.
This is also apparent in Figs.~\ref{viscofig} and \ref{figeffrad},
where the spurious viscosity and the effective hydrodynamic radius
both increase not only with the DNA concentration but also with the colloid 
radius\/.

Fig.~\ref{fig:collaps} shows the drag force normalized to the
pulling velocity $u$ on a colloid of diameter $R_C =
1.12~\mu\text{m}$ measured as a function of the DNA-density
$\rhodna$\/. The data point at $\rhodna=0$ is normalized to the
Stokes drag force with the viscosity of water $\eta_w =
10^{-3}\,\text{N}\,\text{s}/(\text{m}^2)$\/. 
Clearly, the additional drag due to the
presence of the DNA in the solution is not linear in $\rhodna$ and
larger as the drag expected from the increased viscosity.
The the errorbars for small velocities in Fig.~\ref{fig:collaps}
are large due to the error in the velocity.

\begin{figure}
\includegraphics[width=\linewidth]{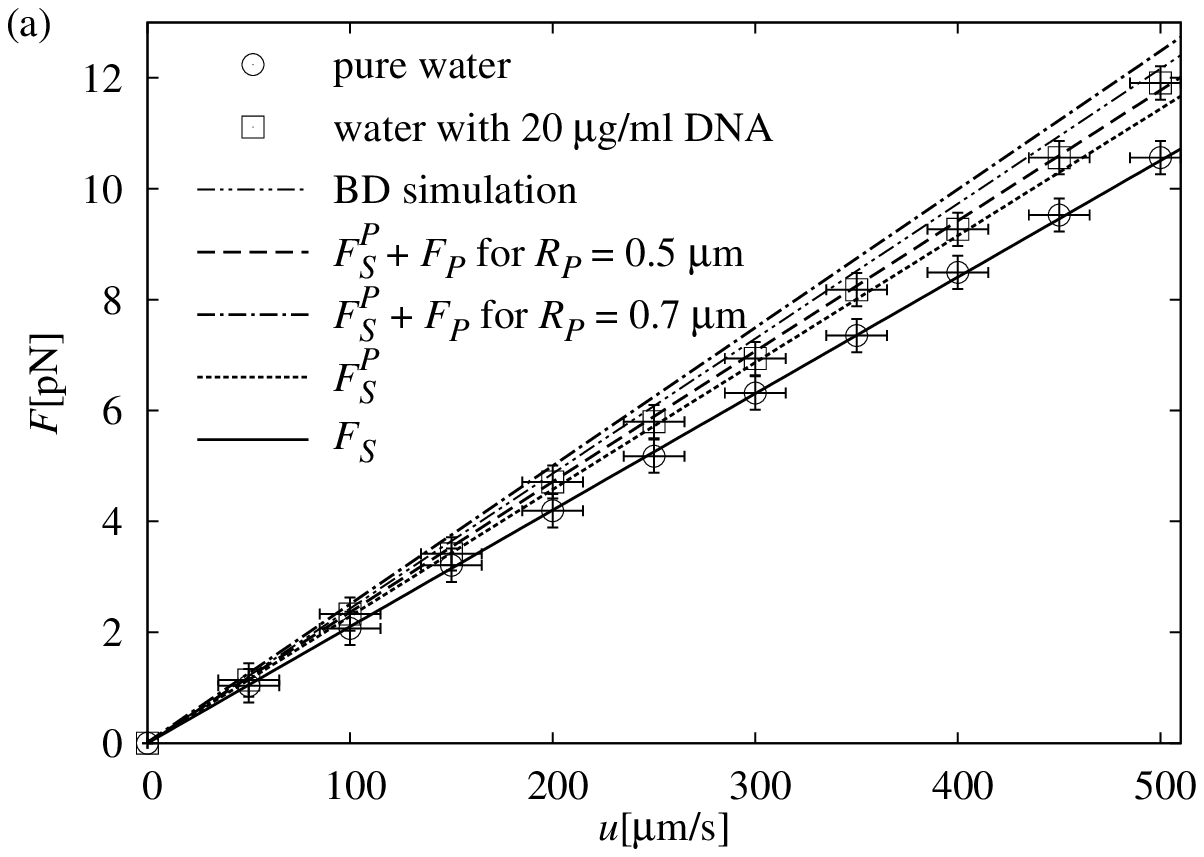}\\
\includegraphics[width=\linewidth]{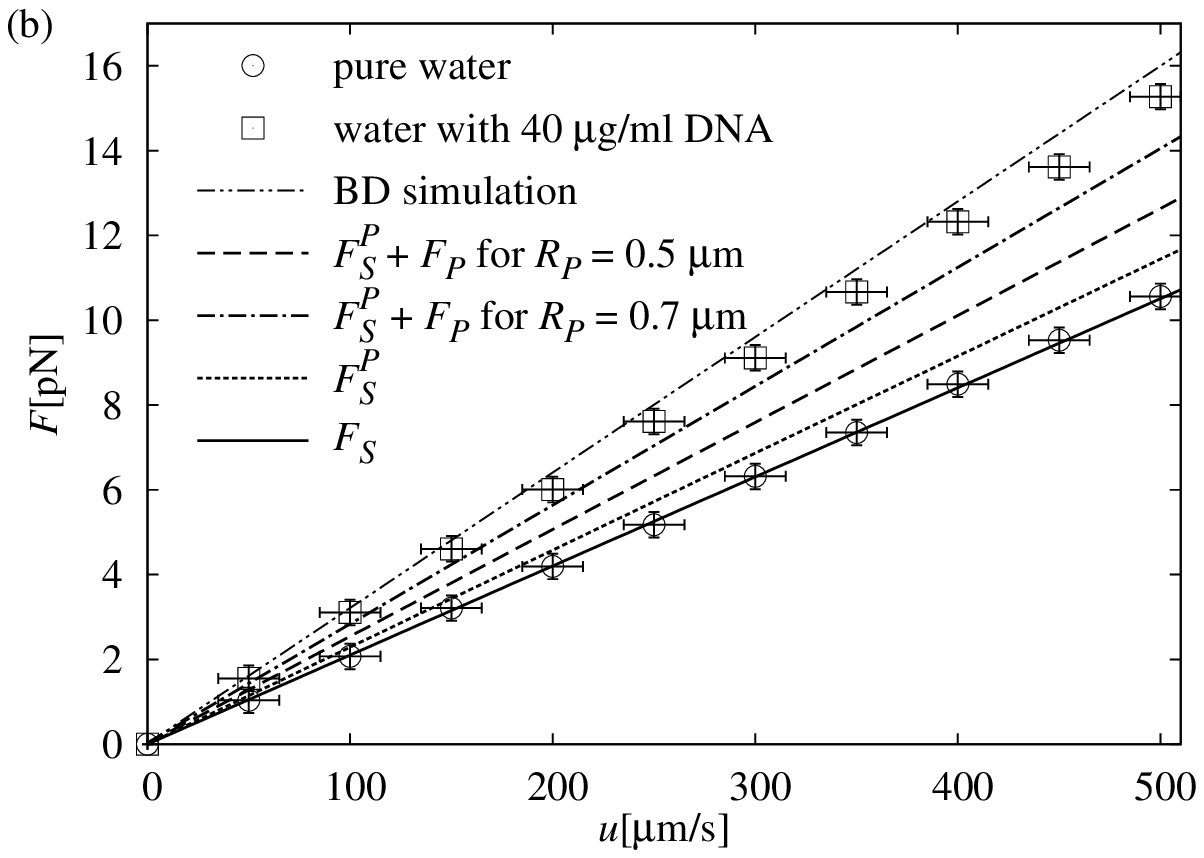}
\caption{\label{fig:force} Drag force $F$ on a colloid of radius
$1.12~\mu\text{m}$ measured in pure water ($\bigcirc$) and in a DNA
solution ($\square$) as a function of the velocity $u$\/. The DNA
concentration in (a) is 20~$\mu$g/ml and in (b) it is
40~$\mu$g/ml\/. The data is compared to the Stokes friction $F_S$
in pure water (full line) and in the DNA solution $F_S^P$ (dotted
line) calculated from the measured viscosity, as well as to $F_S^P$
plus the contribution $F_P$ from the DNA jam in front of the
colloid for $R_P=0.5\,\mu\text{m}$ (dashed line) and for
$R_P=0.7\,\mu\text{m}$ (dashed-doted line)\/. 
Also shown is a fit to BD simulation results between 
$0\mu$m/s and $50\mu$m/s (dash-dot-dotted line)\/.}
\end{figure}

In
Fig.~\ref{fig:force} the measured drag force on the colloid is
compared to the Stokes drag force $F_S^P$ in the DNA solution, see
Eq.~(\ref{eqstokes})\/.
The forces are to a good approximation linear in $u$\/.
The difference between the drag forces for 20~$\mu$g/ml DNA and for
pure water is only significant for velocities larger than
400~$\mu$m/s\/. For 40~$\mu$g/ml DNA we can observe a significant
difference already at velocities larger than 100~$\mu$m/s\/. The
experimental resolution limits the measurment of the DNA induced
drag force to relatively large velocities and to large DNA
concentrations. However, Fig.~\ref{fig:force}(b) clearly shows that
the measured drag forces cannot be explained simply by an increased
viscosity $\eta_\text{DNA}$ of the solution.

\section{Drift-diffusion model}

\begin{figure}
\includegraphics[width=\linewidth]{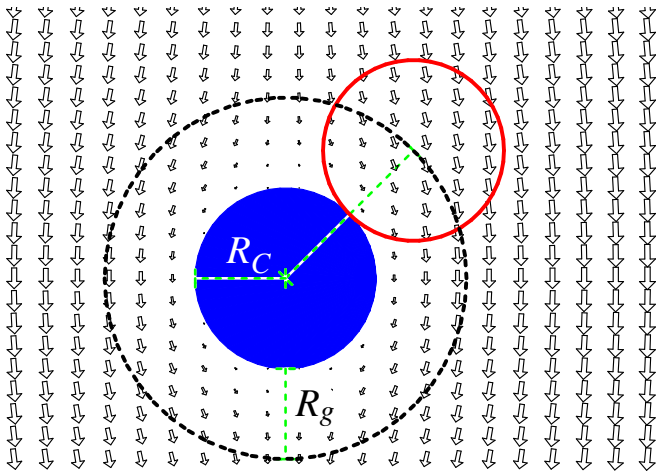}
\caption{\label{fig:sphereflow} Flow field $\vct{v}(\vct{r})$ 
(arrows) around a moving colloid (full
circle) with radius $R_C$ in the frame of reference comoving with
the colloid. Particles (e.g., DNA, open circle) with radius
$R_g$ can approach the colloid only up to a distance $R_C+R_g$
(dashed circle)\/. The flow field has a component normal to this
circle which leads to an accumulation of particles infront and a
depletion of particles in the back of the colloid. }
\end{figure}

That the drag force on the colloid cannot be explained simply by
the increased viscosity of the DNA solution and the dynamics of the
DNA in the vicinity of the moving colloid has to be taken into
account, which is gouverned 
by the interplay of direct intermolecular interactions,
hydrodynamics, and the internal degrees of freedom of a polymer
chain. It has been shown, that an additional drag force due to the
rearrangement of solute particles in the vicinity of a dragged
colloid can be already obtained in a simple drift diffusion (DD) model
\cite{rauscher07b}\/. Here we employ the same model
to calculate this additional drag force $F_P$ and compare it to the
experimental values. Within this model, the DNA molecules are
idealized as mutually non-interacting particles with a finite hard
core interaction radius with the colloid. The origin of this
additional force is illustrated in Fig.~\ref{fig:sphereflow}\/. Due
to the repulsion of the DNA coils and the colloid, the center of
mass of the DNA coils can approach the colloids surface only up to a
distance of roughly $R_g$, i.e., the radius of gyration, which
creates a forbidden zone for the DNA. The solvent molecules,
however, are much smaller and enter this zone, such that the
solvent flow field has a component normal to the surface of the
forbidden zone pointing inwards infront of the colloid and outwards
behind the colloid. DNA coils advected with the solvent will
therefore accumulate infront of the colloid and their density will
be reduced in its back. This inhomogeneous DNA distribution will
lead to an inhomogeneous osmotic pressure and therefore to a force
on the colloid.

We will quantify the rearrangement of DNA molecules by calculating the
average concentration $C(\vct{r})$ (in units of molecules per volume) near
the colloid, which, 
in the simple drift-diffusion (DD) model described in
\cite{squires05b,rauscher07b},
is given by the stationary solution of the Smoluchowski equation
\begin{equation}
\label{eq:ideal}
\vct{v}\cdot\grad C =D\,\Delta  C
\end{equation}
in a frame of reference comoving with the colloid, with the solvent
velocity field $\vct{v}$ and the DNA (zero concentration) diffusion
constant $D$\/. In contrast to Ref.~\cite{squires05b}, $\vct{v}(\vct{r})$
is not uniform but it is the solution of Stokes' equation for a sphere of
radius $R_C$ translating with velocity $\vct{u}$ through a resting solvent
($\vct{v}\to -\vct{u}$ for $|\vct{r}|\to \infty$) (see, e.g.,
Ref.~\cite{landaulifschitzVI} for details),
\begin{multline}
\label{eq:sphereflow}
\vct{v}(\vct{r}) +\vct{u} =\\
\frac{3\,R_C}{4\,r}\,\left[
\left(1+\frac{R_C^2}{3\,r^2}\right)\,
\vct{u} + 
\left(1-\frac{R_C^2}{r^2}\right)\,\vct{\hat{r}}\,(\vct{\hat{r}}\cdot\vct{u})
\right],
\end{multline} 
with the unit vector $\vct{\hat{r}}=\vct{r}/|\vct{r}|$\/.
As we model the interaction potential between the colloid and the
polymers as a hard-sphere potential, the centers of the polymers
are excluded from a sphere of radius $R_C+R_g$ around the center of
the colloid, see Fig.~\ref{fig:sphereflow}\/. Therefore, the
boundary condition for Eq.~\eqref{eq:ideal} on this sphere is 
\begin{equation}
\label{eq:BC}
\left.\left(\uvct{e}_r\cdot\vct{j}\right)\right|_{r=R_C+R_g}=0,
\end{equation} 
i.e., the DNA current $\vct{j}=\vct{v}\, C-D\,\grad C$ normal
to the sphere's surface has to vanish. Far from the colloid, the DNA
density should be constant, i.e., $ C(\vct{r})\to C_0$ for
$|\vct{r}|\to\infty$\/. As the mutual interaction between the DNA
molecules is neglected, the local pressure on the colloid surface
can be calculated from the ideal gas law $p=k_BT\, C$\/.
Integrating the local pressure over the surface yields the force
$F_P$ on the colloid.

The solution of Eq.~\eqref{eq:ideal} depends only on the
dimensionless velocity $u^*=u \frac{R_C+R_g}{D}$ (the Peclet
number) of the colloid and the size ratio of the involved particles
$R^*=R_C/(R_C+R_g)$\/. For $R_g/R_C\to 0$, i.e., for $R^*\to 1$, the
surface of the forbidden zone coincides with the colloid surface
and the DNA behaves like a solvent molecule. In this limit the
solution of Eq.~\eqref{eq:ideal} is $ C(\vct{r})= C_0$, the DNA
molecules do not accumulate infront of the colloid and therefore the
additional drag force $F_P$ vanishes.

\begin{figure}
\includegraphics[width=\linewidth]{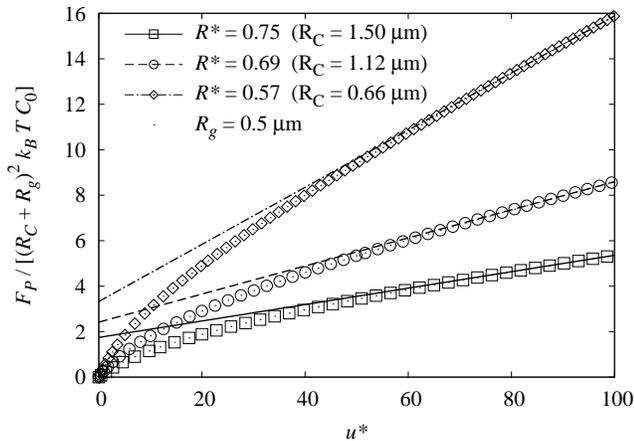}
\caption{Numerically calculated drag force for the radii used in
the experiment as a function of the
Peclet number $u^*$ (symbols) and the affine fits to the data for
$80 \le u^* \le 100$\/.
\label{fig:forcefit}}
\end{figure}

With a
hydrodynamic radius of $R_H=0.33\,\mu$m the Stokes-Einstein
relation leads to a diffusion constant $D\approx 6\times
10^{-13}\,{\text{m}^2}/{s}$\/. Therefore, the smallest
velocities in the experiments ($u=50\,\mu\text{m}/\text{s}$)
correspond to Peclet numbers $u^*$ larger than 100 (depending on
the colloid radius)\/. While for small $u^*$ Eq.~\eqref{eq:ideal}
can be solved analytically in linear order in the Peclet number, we
have to use numerical methods for the experimental velocities.
To this end, we expand the density field
$ C(\vct{r})$ in spherical harmonics up to order $N$ and obtain a
system of $N+1$ ordinary differential equations for the
$|\vct{r}|$-dependent expansion coefficients which we solved numerically with
AUTO~2000 \footnote{http://sourceforge.net/projects/auto2000/} 
(for details, see Ref.~\cite{rauscher07b})\/. 
For large
$u^*$ a fine numerical discretization (large $N$) is needed since
the thickness of the region in front of the colloid in which the
DNA density is enhanced decreases with $u^*$ while the density in
this region increases with $u^*$\/. We therefore use
$N=100$, which allows us to calculate reliable solutions up to
$u^*\approx 100$\/. For such high velocities the drag force $F_P$ is well
approximated by an affine function (see Fig.~\ref{fig:forcefit} and
also Ref.~\cite{squires05b})
\begin{equation}
F_P(R^*,u^*)/[(R_C+R_g)^2\,k_BT\, C_0]=\alpha(R^*)+\beta(R^*)\,u^*.
\end{equation} 
which we use to extrapolate to
velocities larger than $u^*=100$, see Fig.~\ref{fig:forcefit}\/.
The coefficients for fits to the numerical data in the range $80
\le u^* \le 100$ for the experimentally relevant values
$R_C=1.50,\,1.12,\,0.66$~$\mu$m and $R_g=0.5$~$\mu$m are given in 
Table~\ref{tab:force}\/. In order to
calculate $F_P$ also for other values of $R^*$ and to test the
sensitivity of our result to variation of $R_g$ we use the linear
interpolation of the coefficients $\alpha(R^*)$ and $\beta(R^*)$ 
\begin{subequations}
\begin{eqnarray}
\alpha(x) &=& 8.146   - 8.427\,x,\\
\beta(x) &=& 0.411 - 0.502\,x.
\end{eqnarray}
\end{subequations}



\begin{table}[t]
  \begin{ruledtabular}
    \begin{tabular}{cccc}
      $R_C$ [$\mu$m]&$R^*=\frac{R_C}{R_C+R_p}$   
		&$\alpha(R^*)$ & $\beta(R^*)$
		\\\hline
      0.66 & 0.57 
		&  3.31 & 0.126 \\
1.12& 0.69 
&2.43 &0.063\\
1.50& 0.75 
& 1.76 & 0.036\\
    \end{tabular}
  \end{ruledtabular}
  \caption{
  \label{tab:force}
  Fitting coefficients for the extrapolation of the force
  $F_P$ to large velocities (Peclet numbers) $u^*$ for the three
  colloid sizes used in the experiments. The radius of gyration is
  assumed to be $R_g=0.5\,\mu$m\/.}
\end{table}


For the ideal gas law which we use to calculate the local osmotic
pressure on the colloid surface we need the number density
$C_0$ rather than the mass density $\rhodna$ as used in
the experiments. With the molecular weight $M=5.23\times
10^{-11}\,\mu$g we get $C_0 \,[\text{m}^{-3}] = 1.91\times
10^{16}\,\rhodna\,[\mu\text{g}/\text{ml}]$\/.


Fig.~\ref{fig:force} compares the forces on a colloid of size
$R_C=1.12\,\mu$m predicted by the DD model to the experimental data.
Although the difference of the force at a concentration of $\rhodna =
20\,\mu\text{g}/\text{ml}$ in pure water (i.e., only the Stokes force) is
hardly significant and within the experimental error explained by the
enhanced viscosity of the DNA solution (Fig.~\ref{fig:force}(a)), theory
and experiment are in better correspondence (in particular for larger
velocities) if one takes into account the force $F_P$ for a DNA radius of
$R_g=0.5\,\mu$m\/.  For higher DNA concentrations (e.g.,  $\rhodna
=40\,\mu\text{g}/\text{ml}$ in Fig.~\ref{fig:force}(b)), where the
measured forces cannot be explained by the increased viscosity, $F_P$
calculated from the DD model for $R_g=0.5\,\mu$m is too small. A
reasonable fit to the data can be obtained for $R_g=0.7\,\mu$m, however,
for this large value the forces predicted for the lower DNA concentrations
(see, e.g., Fig.~\ref{fig:force}(a)) are too large.

The experimentally measured drag force as a function of the DNA
concentration as shown in Fig.~\ref{fig:collaps} clearly shows a
nonlinearity indicating that interactions among the DNA molecules
are relevant for concentrations larger than $\rhodna =
20\,\mu\text{g}/\text{ml}$\/. The DD model
presented in this section does not take these into account. 
Therefore, Brownian dynamics simulations are performed as presented in the
following section.

\section{Brownian Dynamics Simulations}

\begin{figure}
\includegraphics[width=\linewidth]{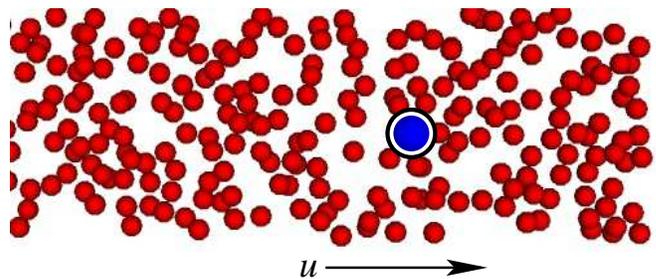}
\caption{\label{fig:simsetup}A cut through a part of the simulated system
(28 $\mu$m $\times$ 10 $\mu$m $\times$ 1.4 $\mu$m\/. Polymers are shown in
red and the colloid in blue. The arrow indicates the direction of motion
of the colloid.  Even though the
density map and the density profiles
shown in Fig.~\protect\ref{fig:density} clearly show an accumulation of
polymers in front of the colloid and a depletion region behind it, this
cannot be seen from a single snapshot of the system.}
\end{figure}

We simulate the experiments with colloids of radius $R_C=1.12\,\mu$m using
a Brownian Dynamics (BD) method -- also called Langevin Dynamics
\cite{bib:gunsteren81}\/. The polymers and the colloid are modeled as hard
spheres with their respective radii. For the polymers we use the radius of
gyration $R_g=0.5\,\mu$m\/.  As the polymers drag along most of the water
contained in their volume, the polymers are assigned the mass of water
contained in a sphere with radius $R_g$\/. The number of polymers in the
system is chosen such that the number density (rather than the mass
density) of the experiments is reproduced. We use a rectangular simulation
volume of $100\times 20 \times 20\,\mu\text{m}^3$
with periodic boundary
conditions in all three directions.  For a polymer concentration of
40~$\mu$g/ml this corresponds to 30592 
polymers.

The colloid is trapped in a moving parabolic potential $V(r)=\frac{1}{2}\,
a\, r^2$, mimicking the optical tweezer.  The potential has a spring
constant of $a=7.5\times 10^{-5}$pN/nm, which gives a better signal to
noise ratio than the experimental value of $8.5\times 10^{-2}$pN/nm\/.
Figure~\ref{fig:simsetup} shows a snapshot of our simulation setup.

In BD, two most important aspects of hydrodynamics felt by the suspended
particles are taken into account, namely the Stokes friction and the
Brownian motion. Correspondingly, this is done by adding to a molecular
dynamics simulation two additional forces.  The Langevin equation
describes the motion a Brownian particle with radius $R$ at position
$\vct{r}(t)$ as
\begin{equation}
m\,\ddot{\vct{r}}(t) =6\,\pi\,\eta\,R\,\dot{\vct{r}}(t)
+\vct{F}_{\text{rand}}(t)
+\vct{F}_{\text{ext}}(\vct{r},t),
\end{equation}
where the first term models the Stokes friction in a solvent of viscosity
$\eta$, $\vct{F}_{\text{ext}}(\vct{r},t)$ is the sum of all external
forces like gravity, forces excerted by other suspended particles, and,
for the colloid, the optical trap. $\vct{F}_{\text{rand}}(t)$ describes
the thermal noise which gives rise to the Brownian motion. The random
force on different particles is assumed to be uncorrelated, as well as the
force on the same particle at different times.  It is further assumed to
be Gaussian with zero mean. The mean square deviation of the Gaussian
(i.e., the amplitude of the correlator) is given by the
fluctuation-dissipation theorem as
\begin{equation}
  \langle| \vct{F}_{\text{rand}}|^2 \rangle = 12\,\pi\,\eta\,R\,k_B\, T .
\end{equation}
In order to reduce computation time, physical quantities are rescaled: the
simulation is carried out at a lower temperature. To compensate for this,
the viscosity of the fluid as well as all energies are scaled by the same
factor. In the present simulations, temperature is scaled down by a factor
of 37500\/. This scaling leaves the diffusion constant as well as the
relative importance of diffusion and motion caused by external forces
unchanged. However, it allows  for a much larger time step (in this case
60$\mu$s) \cite{bib:jens-hecht-ihle-herrmann:2005}\/.

\begin{figure}
\includegraphics[width=\linewidth]{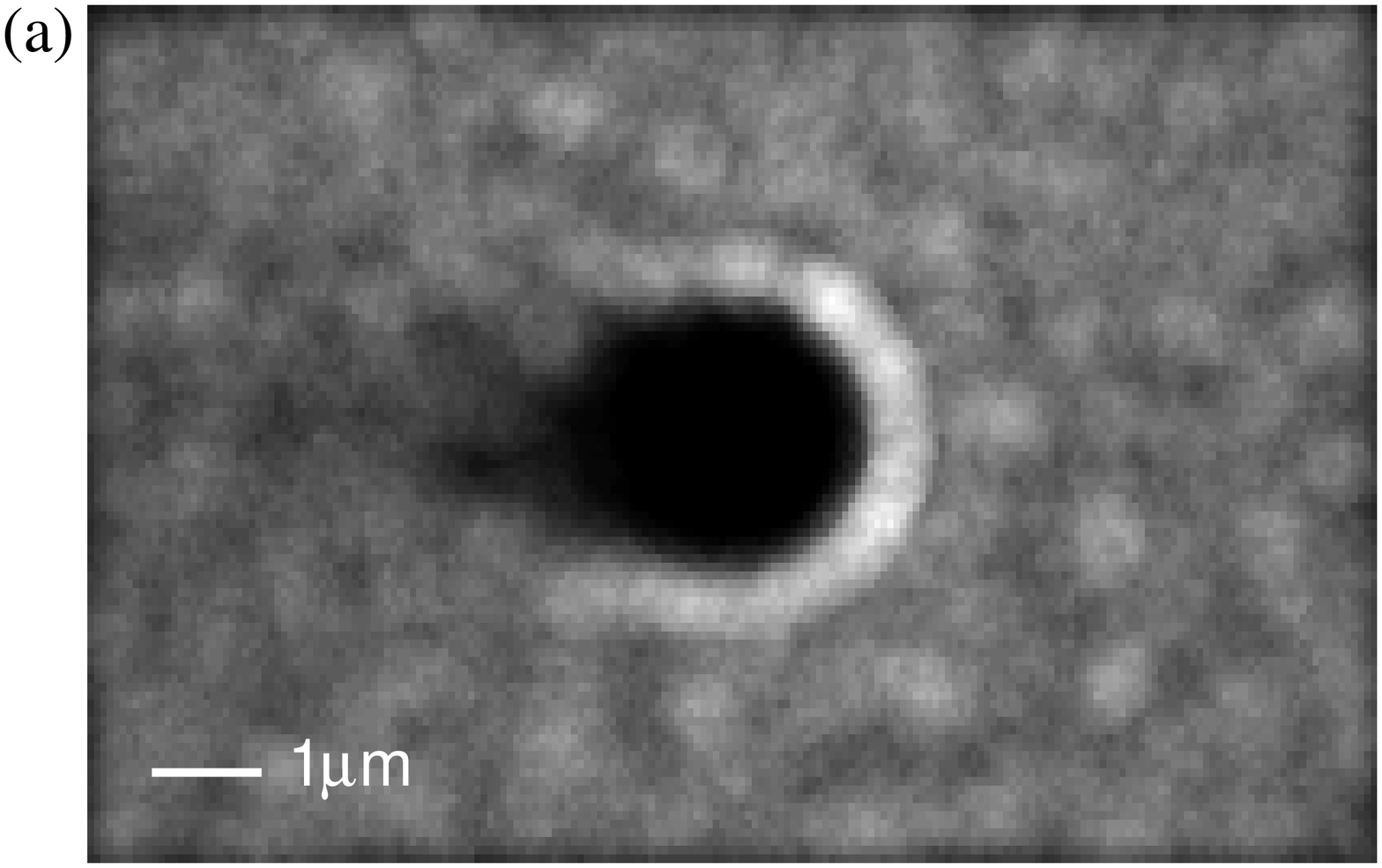}\\
\includegraphics[width=\linewidth]{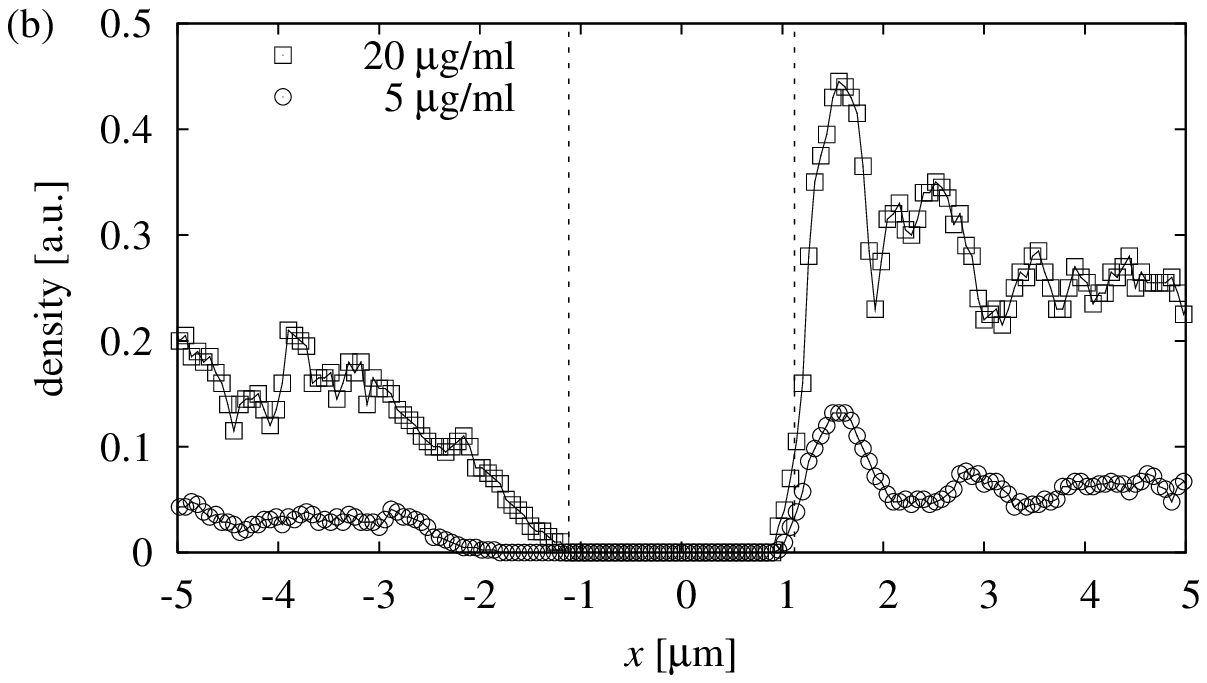}
\caption{\label{fig:density}(a) Polymer density around the colloid
averaged over 600 snapshots of the system at a concentration of
20~$\mu$g/ml and $u=50\,\mu$m/s\/. Lighter colors denote higher polymer
densities. (b) Normalized average polymer density on a line through the
colloid center in the direction of motion for concentrations of
$5\,\mu$g/ml (corcles) and $20\,\mu$g/ml (boxes) and velocity $u=50\,\mu$m/s\/.  Polymers
accumulate infront of the colloid and the concentration in the back is
reduced due to the finite Peclet number of the polymers. Lines
connecting datapoints are guides to the eye.  Also visible are
density oscillations infront of the colloid, which are characteristic for
hard sphere systems. The vertical lines indicate the size of the
colloid ($\pm 1.12\,\mu$m)\/. }
\end{figure}

From the simulation data, it is possible to measure the effective polymer
concentration around the dragged colloid. To accomplish this, about 600
snapshots of the simulation are taken: Each snapshot is moved, so
that the position of the colloid coincides, and the probability for
a certain space 
to be occupied by a polymer is calculated by averaging over all snapshots
(200 x 100 bins are used). The picture in
Fig.~\ref{fig:density}a depicts a typical example of a density map for a
concentration of 20~$\mu$g/ml and $u=50\,\mu$m/s\/. It can be observed
that there is a region of high polymer concentration infront and at the
sides of the colloid whereas the region behind the colloid is not yet
completely refilled by the polymers.  Figure \ref{fig:density}b shows the
normalized average density of polymers in the direction of motion for
concentrations of 5$\mu$g/ml and 20 $\mu$g/ml and $v$=50$\mu$m/s. Infront
of the colloid, a sharp peak can be observed. In addition, on observes
density oscillations which are typical for hard sphere suspensions. For
high polymer concentrations, the probability is close to 1 to find a
polymer infront of the colloid. The region behind the colloid is almost
clear of polymers, because polymers did not yet have enough time to relax
into this region again.

BD is widely used to simulate suspensions (e.g.
\cite{reichhardt028301,reichhardt041405,reichhardt108301}) because it is
well understood, not difficult to implement, and needs much less
computational resources than a full simulation of the fluid.  However,
this simulation method does not resolve more complicated hydrodynamic
phenomena, in particular long-ranged hydrodynamic interactions among the
polymer particles and in particular between the polymers and the colloid.
As a result both, the friction and the thermal force excerted on a
particle are independent of the position and velocity of the other
particles.  In the case of a polymer trapped infront of the moving colloid
(i.e., if it is dragged along by the colloid), this leads to a
considerable overestimation of the friction felt by the colloid: in a BD
simulation it feels the friction on itself and the dragged polymer,
whereas in reality the friction on the colloid would be significantly
reduced because it is in the polymer's slipstream.  In addition, as argued
in Ref.~\cite{rauscher07b}, the hydrodynamic interaction of the colloid
with the polymers is of importance. The solvent molecules flow around the
colloid and the resulting flow field advects the polymers around the
colloid (see Eq.~\eqref{eq:sphereflow}), therefore significantly reducing
the polymer jam infront of the colloid.  To compensate for both of these
effects we reduced the friction felt by the polymers, i.e., increase their
diffusion constant.  This does not significantly influence the density
distribution outside the interaction region where the polymers are
homogeneously distributed.

\begin{figure}
\includegraphics[width=\linewidth]{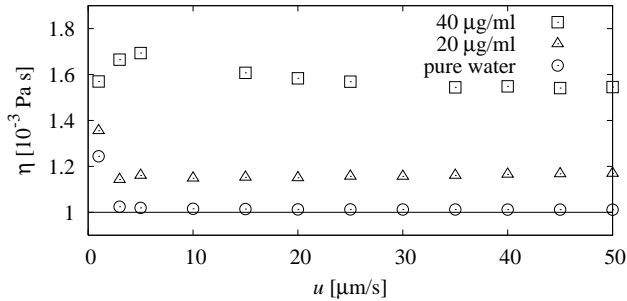}
\caption{\label{img:velocity} Effective viscosity (calculated with
Eq.~\protect\eqref{eqstokes} from drag force measured in the simulations) 
versus drag velocity for different polymer concentrations. For
velocities larger than about $10\,\mu$m/s, the effective 
viscosity is independent of the velocity as expected. For low velocities, 
the drag force increases. 
The horizontal line
indicates the viscosity of the solvent water. }
\end{figure}

Our model reproduces very well the linear relation between drag force and
the drag velocity for different polymer concentrations. As higher drag
velocities require a larger system (even with periodic boundary
conditions) and short numerical timesteps, we are limited to about fifty
micrometers per second by the available computational resources and time.
However, this is not a problem because of the linearity of the drag force
with respect to velocity for higher velocities as observed in the
experiments. Fig.~\ref{fig:force} includes a linear fit to the simulation
data for velocities between $0\mu$m/s and $50\mu$m/s, which agrees well
with the experimental data.

As in the experiments, with Eq.~\eqref{eqstokes} the measured drag force
can be interpreted in terms of an effective viscosity.
Fig.~\ref{img:velocity} shows the dependency of this viscosity on the
velocity for pure water and for polymer concentrations of 20$\mu$g/ml and
40$\mu$g/ml\/. Even for a polymer concentration of 40$\mu$g/ml the
viscosity stays constant for drag velocities greater than 30$\mu$m/s, but
for very low velocities, in all three cases we find that the drag force
decreases less then linearly.  For such low drag velocities the colloid is
very near the trap center and thermal fluctuations about the average
position are large. An energy of $k_{\rm B} T$ corresponds to a
displacement of $0.3\, \mu$m, whereas the displacement calculated from the
drag force in water at $u=1\,\mu$m/s corresponds to a displacement of
$x=0.28\mu$m\/. However, these fluctuations should be roughly symmetric
with respect to the average colloid position within the trap. However, the
force calculated numerically within the framework of the DD model shown in
Fig.~\ref{fig:forcefit} analyzed in terms of an effective viscosity would
also give an increased viscosity for smaller velocities: the slope is
larger for smaller values of $u^*$\/. Therefore we conclude, that the
apparent increase in viscosity as shown in Fig.~\ref{img:velocity} is at
least partially related to the formation of the polymer jam infront of the
colloid. However, the dataset for pure water (i.e., without polymers)
indicates, that at very low velocities, fluctuations do play a role.

For a fixed drag velocity and varying polymer concentrations, the linear
relation between force and concentrations for low polymer concentrations
and the stronger nonlinear increase for higher concentrations can also be
observed as shown in Fig.~\ref{fig:collaps}\/. The increase in drag force
is overestimated as expected if the mobility of the polymers is not
increased (as described above)\/. When the mobility of the polymers is
increased by a factor of 14.3, however, quantitative agreement is achieved
for concentrations up to 40 $\mu$g/ml\/. For even higher concentrations,
the drag force increases stronger than  in the experiments. This is
because at high concentrations, a hard sphere suspension has a stronger
tendency to jam than a real polymer suspension: polymers are deformable
such that, in contrast to hard spheres, they can squeeze past each other.
The simulation results for increased polymer mobility indicate that the
friction felt by a pair of a colloid and a dragged polymer in the
experiment is far less than the sum of the friction forces on a single
polymer and colloid, respectively. 

\section{Discussion}
The experiments clearly show that the drag force on colloids pulled
through a solution of $\lambda$-DNA with an optical tweezer cannot be
explained by the Stokes force for the viscosity of the solution. It is
much higher and increases non-linearly with the DNA-concentration but
approximately linearly with the drag velocity. The resolution of the force
measurment is $\pm 0.15$~pN and therefore the difference between the
forces on the colloid in the solution and in pure water can be measured
only for large concentrations or, in the case of smaller concentrations
only for large velocities. This limits the overlap between the DD theory
and the experiments to the lowest concentration of $\rhodna
=20\,\mu\text{g}/\text{ml}$ used in the experiment. 

At this concentration,
the volume fraction of the solution taken by the DNA coils (based on
$R_g=0.5\,\mu$m) is 0.1 and DNA-DNA interactions can be neglected. The
agreement between theory and experiment is very good. For the highest
concentration used in the experiments $\rhodna =
50\,\mu\text{g}/\text{ml}$ the volume fraction is 0.5 and DNA-DNA
interactions have to  be taken into account. 

In the BD simulations DNA-DNA interactions have been taken into account.
As in the experiments and in the DD theory, the drag force increases
linearly with the velocity, at least  for large velocities. The nonlinear
increase of the drag force with the DNA concentration is also observed in
the BD simulations. However, since the BD simulations could not take into
account hydrodynamic interactions between the colloid and the DNA
molecules, the drag force is significantly overestimated. The hydrodynamic
flow field around the colloid as shown in Fig.~\ref{fig:sphereflow} has a
component normal to the direction of motion of the colloid which
efficiently reduces the number of DNA molecules accumulated infront of the
colloid, while it fills the depletion zone at its back. As demonstrated by
the excellent agreement with the experimental data, this hydrodynamic
interaction between the colloid and the DNA has the same effect as
increasing the DNA mobility, here by a factor of 14.3\/.
This is consistent
with results of BD simulations of soft particles in a background flow,
where the drag force was reduced by a factor of 10 by this hydrodynamic
effect \cite{rauscher07b}\/.

The agreement between theory and experiment for low concentrations but
high velocities is particularly remarkable considering the simplicity of
the models which neglect many aspects of hydrodynamic interactions or
include them in a rather simplistic way (e.g., in terms of an increased
viscosity in the DD model)\/.  The internal structure of the DNA molecules
is neglected in both, the DD model as well as the BD simulations.
Apparently, the deformation of the DNA coils in the accumulation region
infront of the colloid is not significant.  This can be only explained by
the proximity of the stagnation point for the solvent flow which limits
the normal component of the the solvent velocity field. At this point the
shear rate is also smaller than at the colloid's side, where it goes up to
over 700~1/s for $u=500\,\mu\text{m}/\text{s}$\/.  With a relaxation time
on the order of 0.1~s \cite{fang05} this amount to a Weissenberg
number Wi on the order of 70, i.e., much larger than one.
Therefore the polymercoils should be significantly distorted and 
viscoelastic effects are to be expected.

In summary, we present a first direct experimental observation of
jamming-induced drag-enhancement on colloids in polymer solutions.
First theoretical approaches to this problem as well as the
Brownian dynamics simulations presented in this paper contain many
simplifications which need to be addressed in the future.

\begin{acknowledgments}
F.K. J.H., and M.R. acknowledge support by the Deutsche
Forschungsgemeinschaft within the priority program SPP~1164 ``Micro- and
Nanofluidics''. J.H. and R.W. also acknowledge fruitful discussions with
M.~Hecht. The computations were performed at Forschungszentrum J\"{u}lich
and at the Scientific Supercomputing Center in Karlsruhe.
\end{acknowledgments}

\end{document}